\documentclass[aps,prx,twocolumn,floatfix,letterpaper,longbibliography]{revtex4-1}
\usepackage{graphicx}
\usepackage{bm}
\usepackage{amsmath,amsfonts}
\usepackage{amsbsy}
\usepackage[colorlinks=true,linkcolor=blue,urlcolor=blue,citecolor=blue,breaklinks=true]{hyperref}
\usepackage[utf8]{inputenc}
\DeclareUnicodeCharacter{03BC}{\mbox{$\mu$}}
\DeclareUnicodeCharacter{2009}{ }
\usepackage{color}
\usepackage{amssymb}

\usepackage[utf8]{inputenc}
\usepackage{colortbl}

\usepackage{xcolor}

\definecolor{imcolor}{rgb}{0.5,0.,0.5}				
\definecolor{kamcolor}{rgb}{0,0.,0.5}

\begin{document}
\title{Scar states in a system of interacting chiral fermions}
\author{I. Martin}
\author{K. A. Matveev}

\affiliation{Materials Science Division, Argonne National Laboratory,
  Argonne, Illinois 60439, USA}

\date{\today}

\begin{abstract}
We study the nature of many-body eigenstates of a system of interacting chiral spinless fermions on a ring. We find  a coexistence of fermionic and bosonic types of eigenstates in parts of the many-body spectrum. Some bosonic eigenstates, native to the strong interaction limit, persist at intermediate and weak couplings, enabling persistent density oscillations in the system, despite it being far from integrability.

\end{abstract}

\maketitle

\section{Introduction}

According to the  Eigenstate Thermalization Hypothesis (ETH) \cite{deutsch1991quantum, Srednicki1994}  the thermodynamic microcanonical expectation value of a physical observable (a few-body local  operator) of a chaotic system can be obtained by taking a quantum expectation value of that operator in any eigenstate at the corresponding energy. To wit, all energy eigenstates are equally good at encoding thermodynamics. 

ETH was presumed to universally apply to general chaotic (nonintegrable) many-body systems \cite{Alessio2016}.  Recent experimental observation of non-ergodic dynamics in a  non-integrable system of interacting Rydberg atoms \cite{Bernien2017} has called the universality of ETH into question. Since then a number of models have been constructed that clearly demonstrated the  possibility of {\em weak\/} violation of ETH, wherein most of the eigenstates satisfy ETH, but a small subset of states throughout the spectrum  do not \cite{Turner2018a, Moudgalya2018b, Ho2019, Iadecola2019, lin2019exact, Pai2019, Ok2019,  moudgalya2020large, Bull2019, Lee2020, Zhao2020, Bull2020,   Voorden2020, ana2020quantum, Moudgalya2020,    o2020tunnels, Mark2020,  Moudgalya2020, Srivatsa2020}.  These special states typically have  suppressed quantum entanglement, and sometimes are equally spaced in energy, allowing their superpositions to experience perfect revivals. In the latter case, macrosocopic observables can also show oscillations in time, something not expected in ergodic thermalizing systems. These oscillations mirror behavior sometimes encountered in single-particle chaotic billiards: upon quantization, classical unstable periodic trajectories engender sequences of spectrally equidistant eigenstates, with the probability density concentrated around their ``parent" classical trajectories---a phenomenon termed quantum scarring \cite{Heller1984}. Superpositions of these scarred eigenstates can capture periodic motion of wavepackets. This analogy has earned the corresponding many body states the name  Quantum Many-Body Scars (QMBS). Note that not only QMBS are equidistant in energy, but their low entanglement (proximity to simple product states) reinforce the analogy with the semiclassical single-body scars.  

More recently, the term QMBS has come to denote any state, even if it is not a part of an equidistant in energy sequence (``tower") as long as it is atypical, and thus violates ETH. Several not mutually exclusive  mechanisms for QMBS formation have been recognized \cite{serbyn2020quantum}. All of them rely on the existence of exactly (or almost exactly) isolated subspace spanned by a subset of the  eigenstates that share a special property, for example, a higher symmetry. 
Yet, the complete understanding of the origin of QMBS and their effects on macroscopic dynamics are  still lacking.  The fact that some nonintegrable systems show long-lived periodic dynamics raises a question whether these systems are close to some integrable point  \cite{khemani2019signatures}. While certainly not necessary for the existence of generalized QMBS obtained by spectral embedding \cite{Shiraishi2017}, in some cases this remains a viable possibility. 

In this paper we consider  one of the simplest physical systems---chiral interacting spinless fermions confined to move on a circle of finite length $L$---to glean a possible relationship between integrability and scarring. This model is inspired by the physics of a quantum Hall droplet. The limiting cases of zero and infinite interactions are exactly solvable and thus completely integrable. However, for general interactions and fermion dispersion the model is certainly nonintegrable and thus should be expected to satisfy ETH. What makes this model also interesting is that the eigenstates in the two extreme integrable limits are fundamentally different: fermionic for zero interactions, and bosonic for infinite interactions.  
The central result of this work is that energies of the bosonic and  fermionic states at the upper edge of the spectrum (for a fixed value of the total momentum $P=2\pi n/L$, where $n$ is integer) {\em intersect\/} as a function of interaction strength. This immediately implies simultaneous presence of bosonic and fermionic states at intermediate couplings at least in part of the spectrum. Given that the fermionic and bosonic states are qualitatively different, but simultaneously present in the same spectral region, the strong version of ETH, which assumes that {\em all\/} eigenstates are qualitatively the same, is automatically violated.

For infinitely strong repulsive interactions, the anti-ground (i.e., the highest energy) state corresponds to multiple excitations occupying the lowest-momentum bosonic mode, $(b_1^\dagger)^n |0\rangle$.
Tracking the whole sequence for different values of $P$, we find that these states form a tower of states with the energy spacing given approximately by the energy of a single $b_1$ boson, even for non-infinite interaction. 
This state is analogous to the exact scar states of the form  $\sim (Q^\dag)^n |0\rangle$ generated by repeated applications of an operator $Q^\dag$ to some reference eigenstate, previously identified in the Hubbard model \cite{PhysRevB.102.075132, Moudgalya2020} and some spin models \cite{Moudgalya2018, Schecter2019}. 

In addition to the anti-ground state, we find that  descendants of other bosonic states can also survive on the nominally fermionic weak-coupling side.  Similarly, the descendants of the fermionic states show  persistence on the nominally bosonic side, within the same model. This  indicates a distinct mechanism of QMBS, that is governed by proximity to two integrable regimes, with a phase transition separating them.

The rest of the paper is organized as follows. In Seciton 
\ref{sec:hamiltonian} we specify the model, expressing it in two equivalent languages: fermionic and bosonic. In Section \ref{sec:f_vs_b} we evaluate the energies of the ground and anti-ground states in the regimes of the weak and strong interactions.  Surprisingly, we find that the ranges of validity of these expansions overlap for the anti-ground state in a fixed total momentum sector, which leads us to the conclusion that scar states must exist near the upper boundary of the spectrum.
In Section \ref{sec:numerics} we present numerical results that corroborate our analytical perturbative results, but also reveal QMBS in the parts of the spectrum not readily accessible analytically. In Section 
\ref{sec:observe} we describe an experimentally feasible way to detect QMBS. In Section
\ref{sec:discussion} we summarize our findings, their implications for thermalization in quantum chaotic systems and possible future directions.

\section{The model}
\label{sec:hamiltonian}

\begin{figure}
\includegraphics[width=.95\columnwidth]{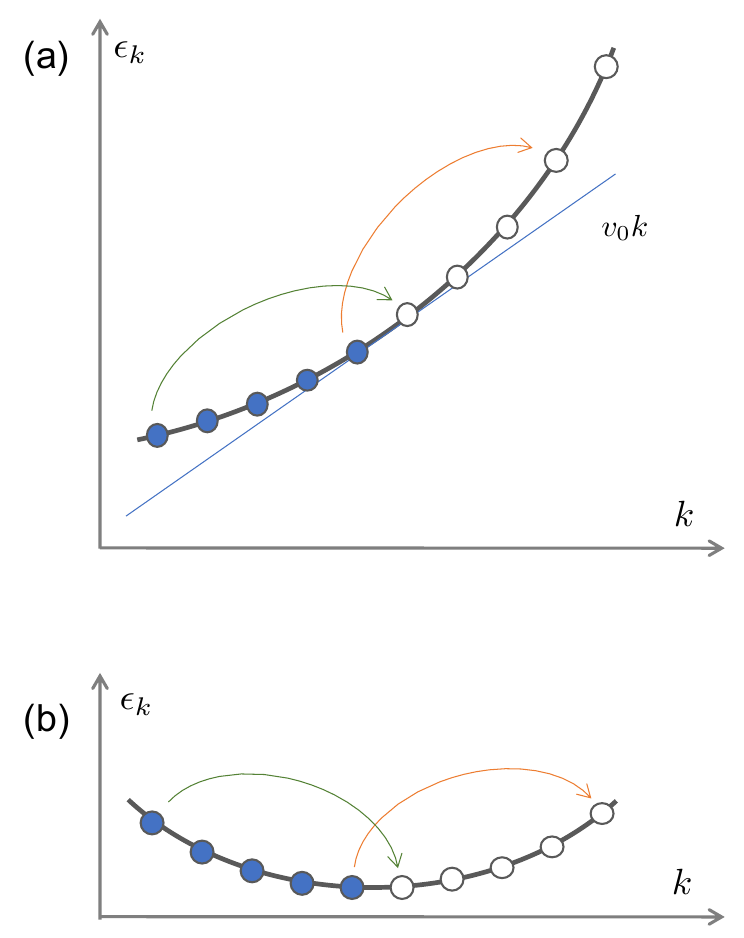}
\caption{\label{fig:dispersion} Chiral fermion dispersion and excitations. (a) Physical spectrum is mostly linear, with a parabolic correction. In the ground state, all states with momentum $k$ below a Fermi momentum are filled, while the rest are empty.
At fixed total momentum $P$, the slope of the dispersion at the Fermi level, $v_0$,  gives only a constant contribution $v_0 P$ to the Hamiltonian, and thus can be ignored. This leads to a simple parabolic dispersion near the Fermi level (panel b).  The orange and green arrows show particle and hole excitations (with momentum 5 in this case).
}
\end{figure}

We consider a model of interacting spinless chiral fermions on a
ring. The noninteracting dispersion $\epsilon_k$ with the dominant
linear part $v_0 k$ and local parabolic curvature is shown in
Fig.~\ref{fig:dispersion}(a). Throughout this paper we assume the
conservation of the total momentum $P$.  In this case the linear part
of $\epsilon_k$ gives a constant contribution $v_0 P$ to the total
energy and does not affect the eigenstates of the system.  We
therefore omit the $v_0 k$ part of the dispersion in the rest of the
paper, see Fig.~\ref{fig:dispersion}(b).  Despite the shift of
energies, the occupancies in the ground state remain the same as in
the Fig.~\ref{fig:dispersion}(a).

As the first step, to make the problem dimensionless, we  take the  circumference $L=2\pi$ and  $\hbar = m =1$. This  corresponds to energy being measured in the units of $4\pi^2 \hbar^2/(mL^2)$.
The momentum of each fermion is given by an integer number,
$k=0, \pm1,\pm2,\ldots$,  and the Hamiltonian of the free
fermions has the form
\begin{equation}
  \label{eq:H_F_f}
  H_F=\sum_k\epsilon_ka_k^\dagger a_k^{},
  \quad
  \epsilon_k=\frac1{2} \left(k-\frac12\right)^2.
\end{equation}
Here $a_k^\dagger$ is the creation operator for a fermion with momentum $k$. 

The simplest eigenstate of $H_F$ corresponds to all single particle states with $k\leq0$ being filled, and those with $k>0$ empty, see Fig.~\ref{fig:dispersion}.  We denote this state $|0\rangle$ and  measure the total momentum $P$ of any state from its value at $|0\rangle$.  It is easy to see that there is only one state in $P = 1$ sector, $a^\dag_1 a_0 |0\rangle$, which corresponds to taking a fermion from the state $k = 0$ and moving it to $k = 1$. In contrast,  there are multiple fermionic states in sectors with $P>1$. Taken as product states (Fock basis), they are eigenstates of $H_F$.  However, interactions  generate mixing among them.

We consider the usual two-body interactions between fermions, which in a translation invariant system are described by the Hamiltonian
\begin{equation}
  \label{eq:H_B_f}
  H_B = \sum_{\substack{k_1 k_2\\ q>0}}
  V(q)a_{k_1+q}^\dagger a_{k_1}^{} a_{k_2}^\dagger
      a_{k_2+q}^{},
\end{equation}
for which $H_B|0\rangle=0$.  To achieve the latter condition we have subtracted from the standard normal-ordered expression an operator that at fixed number of particles amounts to a (infinite) constant.
In addition, in Eq.~(\ref{eq:H_B_f}) we omitted the $1/L=1/2\pi$ prefactor, subsuming it in the  definition of the Fourier transform $V(q)$ of the interaction potential.

In the limit of strong interactions, it is convenient to bosonize the
Hamiltonians (\ref{eq:H_F_f}) and
(\ref{eq:H_B_f}).  Bosonic operators $b_q$ are
expressed in terms of the fermionic ones as \cite{haldane_luttinger_1981}
\begin{equation}
  \label{eq:b_q}
  b_q=\frac{1}{\sqrt q}\sum_k a_k^\dagger a_{k+q}^{},
  \quad
  b_q^\dagger=\frac{1}{\sqrt q}\sum_k a_{k+q}^\dagger a_{k}^{},
\end{equation}
where the momentum $q$ of the boson is a positive integer.
Using Eq.~(\ref{eq:b_q}), one can rewrite the interaction Hamiltonian
(\ref{eq:H_B_f}) as
\begin{equation}
  \label{eq:H_B_b}
  H_B=\sum_{q=1}^\infty \varepsilon_qb_q^\dagger b_q^{},
  \quad
  \varepsilon_q=qV(q).
\end{equation}

The main advantage of bosonization is that  Hamiltonian (\ref{eq:H_B_b}) is quadratic in bosonic variables.  The disadvantage is that in terms of the bosonic variables the Hamiltonian $H_F$ is no longer quadratic \cite{haldane_luttinger_1981},
\begin{equation}
  \label{eq:H_F_phi}
  H_F=\frac{1}{12\pi}\int_0^{2\pi}[\phi'(x)]^3dx,
\end{equation}
where the bosonic field $\phi(x)$ is defined as
\begin{equation}
  \label{eq:phi}
  \phi(x)=-i\sum_{q=1}^\infty \frac{1}{\sqrt q}(b_qe^{iqx}-b_q^\dagger
  e^{-iqx}).
\end{equation}
Substituting this expression into Eq.~(\ref{eq:H_F_phi}), we obtain
\begin{equation}
  \label{eq:H_F_b}
  H_F=\frac12\sum_{q=2}^\infty \sum_{q'=1}^{q-1}
  \sqrt{qq'(q-q')}
  \left(b_q^\dagger b_{q'}b_{q-q'}+ b_{q'}^\dagger b_{q-q'}^\dagger b_q\right).
\end{equation}

In the bosonic representation, the fully occupied Fermi sea $|0\rangle$
corresponds to the bosonic vacuum.  The only
fermionic state $a^\dag_1 a_0 |0\rangle$ in the $P = 1$ sector
corresponds to $b_1^\dag |0\rangle$ state in the bosonic
representation. Despite very different representations, the number of
states in a given sector $P$ does not depend on the basis
\cite{haldane_luttinger_1981}. One can easily verify explicitly that
the expressions (\ref{eq:H_F_f}) and (\ref{eq:H_F_b}) give the same
excitation spectrum for the system of free fermions for a few small
values of the total momentum $P$.

Bosonic basis provides a simple way to count the number of states per sector.  It is given by the number of distinct integer partitions of $P$, i.e., the number of distinct sets of non-negative integers $\{n_q\}$, with $q = 1, \ldots, P$ such that $\sum_q q\, n_q = P$. Physically, $q$ is the momentum of a bosonic mode, and $n_q$ is its occupation number.  For large total momentum, $P\gg 1$, the dimensionality of the symmetry sector scales as $\frac1{4 \sqrt{3}P} \exp\left(\pi\sqrt{\frac{2 P}{3}}\right)$, Ref.~\cite{hardy_asymptotic_1918}. We have been  able  to compute all eigenstates for $P \leq 38$, and some eigenstates for $P$ up to 72, which allowed us  to numerically test  analytical findings.

In this paper we will concern ourselves primarily with unscreened Coulomb interaction between fermions, which we define as  $V(x)\propto \sum _{n = -\infty}^{\infty}1/|x-2\pi n|$ to account for the periodic boundary conditions.
The Fourier transform of this interaction potential has the form
\begin{equation}
  \label{eq:V_q_Coulomb}
  V(q)=-V_0\ln |q|.
\end{equation}
Just like the individual momenta of electrons, for $L = 2\pi$, the relevant momentum transfers $q$  are limited to integers. We have also used the freedom to choose the local part $\delta V(x)\sim\delta(x)$ of electron-electron interaction, which does not affect physics due to the Pauli exclusion principle, such that the associated constant offset in the momentum space results in $V(1) = 0$.

\section{Crossover between the regimes of weak and strong
  interactions}
\label{sec:f_vs_b}

In this section we provide analytical results for some eigenstates in
the intermediate coupling regime based on the weak and strong coupling
expansions. Most importantly, we demonstrate that the ranges of validity
of the two expansions can overlap. The weak coupling expansion
perturbatively modifies fermionic product states, and the strong
coupling expansion does the same to bosonic states. The overlap thus
implies that there is a range of interactions where fermionic and
bosonic types of states coexist.  The states of one type in a region
dominated by those of the other type represent scar states in our
model.

Let us consider the states in the sector with total momentum $P$
relative to the fully occupied Fermi sea state, corresponding to
$P = 0$.  In the case of free fermions, $V_0=0$, described by the
Hamiltonian (\ref{eq:H_F_f}), each state has a number of particle and
hole excitations, with the total momentum $P$.  Due to the convexity
of the kinetic energy, the upper and lower boundaries of the energy
spectrum are achieved for the states
\begin{equation}
  \label{eq:psi_F}
  \psi_F^{(u)}=a_P^\dagger a_0^{}|0\rangle,
  \quad
  \psi_F^{(l)}=a_1^\dagger a_{1-P}^{}|0\rangle.
\end{equation}
They correspond to a single particle and a single hole excitations,
respectively.  (These are the states that are obtained from the fully
occupied Fermi sea via processes shows by green and orange arrows in
Fig.~\ref{fig:dispersion}.) The energies of these states are
\begin{equation}
  \label{eq:E_F}
  E_F^{(u)}=\frac{P(P-1)}{2},
  \quad
  E_F^{(l)}=-\frac{P(P-1)}{2}.
\end{equation}
In the opposite limit of strong interactions, $V_0\to+\infty$, one can
neglect $H_F$ compared to $H_B$, and use the form (\ref{eq:H_B_b}) of
the latter to determine the boundaries of the excitation spectrum.
The states with the highest and lowest energies are
\begin{equation}
  \label{eq:psi_B}
  \psi_B^{(u)}=\frac{1}{\sqrt{P!}}(b_1^\dagger)^P|0\rangle,
  \quad
  \psi_B^{(l)}=b_P^\dagger|0\rangle.
\end{equation}
Their energies are 
\begin{equation}
  \label{eq:E_B}
  E_B^{(u)}=0,
  \quad
  E_B^{(l)}=-V_0 P\ln P.
\end{equation}

As we see, these spectral boundary states, both in the fermionic and
bosonic limits correspond to very simple states.  Next we use
perturbation theory to obtain the shapes of the upper and lower
boundaries of the spectrum in the regimes of weak and strong
interactions.  This will enable us to obtain the scale of the
interaction parameter $V_0$ at which the crossover occurs and to
discuss its nature.

\subsection{Weak interaction regime}
\label{sec:upper_fermion}

We start by evaluating the first order correction to the fermionic
anti-ground state energy $E^{(u)}_F$ in the case of weak interactions.
To this level of approximation the correction to the energy of the
system is obtained by evaluation of the expectation value of the
interaction Hamiltonian (\ref{eq:H_B_f}) in a state with
$n_k = \langle a_k^\dagger a_k^{}\rangle$ taking values 0 or 1.  This
procedure yields
\begin{equation}
  \label{eq:delta_E_F}
  \delta_1 E_F = \sum_{k,k'}n_k (1-n_{k'})V(k-k')\theta(k-k').
\end{equation}
The sum here is performed over all integer $k$ and $k'$.  Technically,
the term with $k=k'$ is excluded from the sum.  This is immaterial
because we have assumed $V(0)=0$, see
Eq.~(\ref{eq:V_q_Coulomb}).

The  state with the highest possible energy at a given momentum $P$
is obtained from the ground state by moving a fermion from $k=0$ to
$k=P$, cf.\ Eq.~(\ref{eq:psi_F}) (also orange arrow in Fig.~\ref{fig:dispersion}).  For this state only the $P$ terms, for which $k=P$ and $k'=0,1,\ldots,P-1$ contribute to
Eq.~(\ref{eq:delta_E_F}), resulting in  
\begin{equation}
  \label{eq:delta_E_F_1}
  \delta_1 E_F^{(u)} = \sum_{q=1}^{P}V(q),
\end{equation}
For the Coulomb interaction $V(q)$ is given by Eq.~(\ref{eq:V_q_Coulomb}), and we find 
\begin{equation}
  \label{eq:delta_E_F_1_Coulomb}
  \delta_1 E_F^{(u)} = -V_0\ln (P!).
\end{equation}
Repeating the same calculation for the lowest energy state $\psi_F^{(l)}$,
obtained by introducing a single hole with momentum $P$, see
Eq.~(\ref{eq:psi_F}) (also green arrow in Fig.~\ref{fig:dispersion}), we find the same result, 
\begin{equation}
  \label{eq:delta_E_F_1_l_Coulomb}
  \delta_1 E_F^{(l)} = -V_0\ln (P!).
\end{equation}
In the most important regime of $P\gg1$, the results
(\ref{eq:delta_E_F_1_Coulomb}) and (\ref{eq:delta_E_F_1_l_Coulomb})
scale as
\begin{equation}
  \label{eq:delta_E_F_1__both_Coulomb}
  \delta_1 E_F^{(u)} =  \delta_1 E_F^{(l)}\simeq -V_0 P(\ln P-1).
\end{equation}

The second order correction to the energy of the anti-ground state at
weak interactions is
\begin{equation}
  \label{eq:delta_E_F_2_general}
  \delta_2 E_F^{(u)}=\sum_{q=1}^{P-1}\sum_{k=0}^{q-1}
  \frac{(V(q)-V(P-k))^2}{q(p-k)}\theta(P-q-k),
\end{equation}
where the unit step function $\theta(x)$ is defined such that
$\theta(0)=0$.  For the Coulomb interaction (\ref{eq:V_q_Coulomb}) we
obtain
\begin{equation}
  \label{eq:delta_E_F_2_Coulomb}
  \delta_2 E_F^{(u)}=c(P)V_0^2,
\end{equation}
where
\begin{equation}
  \label{eq:c(p)}
  c(P)=\sum_{q=1}^{P-1}\sum_{k=0}^{q-1}
  \frac{\ln^2\frac{q}{P-k}}{q(P-k)}\theta(P-q-k).
\end{equation}
Repeating the evaluation for the single-hole state $\psi_F^{(l)}$, one
obtains
\begin{equation}
  \label{eq:delta_E_F_2_l_Coulomb}
  \delta_2 E_F^{(l)}=-c(P)V_0^2.
\end{equation}
At large momentum $c(P)$ defined by Eq.~(\ref{eq:c(p)}) approaches the
limit $c(\infty)=7\pi^4/360$.

\subsection{Strong interaction regime}
\label{sec:upper_boson}

We now study the energies of the ground and anti-ground states in the
regime of strong interactions.  To first approximation one can neglect
the term $H_F$ that does not contain the large parameter $V_0$ and
obtain the energies given by Eq.~(\ref{eq:E_B}).  Next we find the
leading order correction to the ground state energy $E_B^{(l)}$ in the
perturbation (\ref{eq:H_F_b}).  Clearly, the first-order correction
$\langle \psi_B^{(l)}|H_F|\psi_B^{(l)}\rangle=0$.  The second-order
correction to the energy is
\begin{equation}
  \label{eq:delta_E_B^l}
  \delta E_B^{(l)}
  =\frac{1}{2}
  \sum_{q=1}^{P-1}\frac{Pq(P-q)}
  {\varepsilon_P-\varepsilon_q-\varepsilon_{P-q}}.
\end{equation}
Substituting the expression $\varepsilon_q=-V_0q\ln q$, we obtain
\begin{equation}
  \label{eq:delta_E_B^l_result}
  \delta E_B^{(l)} = -\gamma(P)\frac{P^3}{V_0},
\end{equation}
where
\begin{equation}
  \label{eq:gamma}
  \gamma(P)
  =\frac{1}{2P^2}
  \sum_{q=1}^{P-1}\frac{q(P-q)}{q\ln\frac{P}{q}+(P-q)\ln\frac{P}{P-q}}.
\end{equation}
At large $P$ the function $\gamma(P)$ approaches a finite value,
$\gamma(\infty) \approx 0.15743$.

To first approximation the energy of the anti-ground state $E_B^{(u)}$
given by Eq.~(\ref{eq:E_B}) vanishes.  The leading order expression
for $E_B^{(u)}$ is obtained in the second order perturbation theory in
the operator $H_F$ given by Eq.~(\ref{eq:H_F_b}).  An important
difference between the wavefunctions of the ground and anti-ground
states is that the latter has a single boson state with the large
occupation number $P$, see Eq.~(\ref{eq:psi_B}).  To properly account
for the large occupation of bosonic states involved in the
perturbation theory, we introduce $|n_1,n_2,n_3,\ldots\rangle$ as the
properly normalized state in which for $q=1,2,3,\ldots$ the bosonic
modes with momentum $q$ are occupied by $n_q$ bosons.  At
$V_0=\infty$, the anti-ground state $\psi_B^{(u)}$ is then written as
$|P,0,0,\ldots\rangle$.  To first order in $H_F$, it couples only to
the state $|P-2,1,0,\ldots\rangle$, with the matrix element
\begin{equation}
  \label{eq:t_12}
  t_{12}=\langle P-2,1,0,\ldots|H_F|P,0,0,\ldots\rangle
  =\sqrt{\frac{P(P-1)}{2}},
\end{equation}
see Eq.~(\ref{eq:H_F_b}). The first order correction to wavefunction
results in the second-order correction to the energy,
$t_{12}^2/(-\varepsilon_2)$, which in the case of Coulomb interaction
takes the form
\begin{equation}
  \label{eq:E_B^u_second-order}
  E_B^{(u)}=\frac{P(P-1)}{4\ln2\,V_0}.
\end{equation}
To determine the range of applicability of this result, we next find
the first subleading correction to $E_B^{(u)}$.

In second order in $H_F$ for the wave function, two additional states become  coupled to
$|P,0,0,\ldots\rangle$.  First, one can transfer two more bosons from
mode 1 to mode 2, yielding $|P-4,2,0,\ldots\rangle$.  The
corresponding matrix element is
\begin{eqnarray}
  \label{eq:t_12-tilde}
  \tilde t_{12}&=&\langle P-4,2,0,\ldots|H_F|P-2,1,0,\ldots\rangle
\nonumber\\
   &=&\sqrt{(P-2)(P-3)}.
\end{eqnarray}
Second, a boson can be transferred from mode 2 to mode 3, while also
removing one boson from mode 1.  The corresponding matrix element is
\begin{eqnarray}
  \label{eq:t_23}
  t_{23}&=&\langle P-3,0,1,\ldots|H_F|P-2,1,0,\ldots\rangle
\nonumber\\
   &=&\sqrt{6(P-2)}.
\end{eqnarray}
Using the above expressions, the energy up to the fourth order has the form
\[
  E_B^{(u)}=-\frac{t_{12}^2}{\varepsilon_2}+
  \frac{t_{12}^2 \left(2 t_{12}^2-\tilde t_{12}^2\right)}{2\varepsilon
   _2^3}
  -\frac{t_{12}^2 t_{23}^2}{\varepsilon_2^2 \varepsilon_3},
\]
which for Coulomb interaction results in
\begin{equation}
  \label{eq:E_B^u_fourth-order}
  E_B^{(u)}=\frac{P(P-1)}{4\ln2\,V_0}
  +
  \left[
    \frac{P-2}{4(\ln2)^2\ln3}
    -\frac{2P-3}{16(\ln2)^3}
  \right]
  \frac{P(P-1)}{V_0^3}.
\end{equation}
The second term here is small compared to the first one as long as
$V_0\gg\sqrt{P}$.

To gain better insight into the nature of the anti-ground state at
strong interactions, let us consider the limiting procedure in
which $P\to\infty$, while $V_0/P$ remains finite.  In this
limit one can simplify Eqs.~(\ref{eq:t_12}) and (\ref{eq:t_12-tilde})
and approximate $t_{12}=P/\sqrt2$ and $\tilde t_{12}=P$.  This
approximation can be understood as replacing
$(b_2^\dagger b_1^{} b_1^{}+ b_1^{\dagger} b_1^{\dagger} b_2^{})\to
P(b_2^\dagger+b_2^{})$ in the perturbation operator (\ref{eq:H_F_b}).
The other terms in Eq.~(\ref{eq:H_F_b}) contain at most one operator
$b_1$ or $b_1^\dagger$, which means that they are proportional to at
most $\sqrt P$ and can be neglected in comparison.  This corresponds
to replacing the full Hamiltonian $H_B+H_F$ with the effective
Hamiltonian 
\begin{equation}
  \label{eq:H_eff}
  H_{\rm eff}=\varepsilon_2 [b_2^\dagger b_2^{} +\lambda(b_2^\dagger + b_2^{})],
\end{equation}
where
\begin{equation}
  \label{eq:lambda}
  \lambda=\frac{P}{\sqrt2\, \varepsilon_2}.
\end{equation}
The Hamiltonian $H_{\rm eff}$ is easily diagonalized,
\begin{equation}
  \label{eq:H_eff_diagonalized}
  H_{\rm eff}={\varepsilon_2} b^\dagger b -\lambda^2{\varepsilon_2}.
\end{equation}
Here the new bosonic operator $b$ is defined by
\begin{equation}
  \label{eq:b}
  b=b_2+\lambda.
\end{equation}
Physically, $\lambda$ is the constant displacement of mode 2 induced by the bosonic ``condensate" in mode 1. The corresponding vacuum energy of mode 2, given by the constant term in Eq.~(\ref{eq:H_eff_diagonalized}), becomes the energy of the anti-ground state
that we seek,
\begin{equation}
  \label{eq:E_B^u}
  E_B^{(u)}=-\lambda^2{\varepsilon_2}=\frac{P^2}{4\ln 2\,V_0},
\end{equation}
At $P\to\infty$, this expression is indeed identical to
the first term of Eq.~(\ref{eq:E_B^u_fourth-order}), but unlike the
latter, our result (\ref{eq:E_B^u}) does not rely on the perturbative
expansion of $E_B^{(u)}$ in the coupling matrix element (\ref{eq:t_12}).

The applicability of this approach is limited by the condition that
the terms $b_2^\dagger b_1^{} b_1^{}$ and
$ b_1^{\dagger} b_1^{\dagger} b_2^{}$ in the perturbation
(\ref{eq:H_F_b}) are dominant.  This holds as long as the occupation
of the first bosonic state $\langle b_1^\dagger b_1^{}\rangle\sim P$
is much larger that that of any other state.  Given the form of the
Hamiltonian (\ref{eq:H_eff}), the second most occupied bosonic state
has momentum $q=2$.  Its occupation
$\langle b_2^\dagger b_2^{}\rangle$ in the vacuum state of the
operator $b$ is
\begin{equation}
  \label{eq:n_2}
  \langle b_2^\dagger b_2^{}\rangle=\lambda^2
  =\frac{1}{8\ln^22}\left(\frac{P}{V_0}\right)^2,
\end{equation}
where we used Eq. (\ref{eq:b}).  Thus the condition
$\langle b_2^\dagger b_2^{}\rangle\ll \langle b_1^\dagger
b_1^{}\rangle\sim P$ is satisfied as long as $V_0\gg\sqrt P$.  This
condition coincides with the one obtained earlier from the
perturbation theory.

Before closing this section, we compute the overlap of the vacuum states of the operators $b$ and $b_2$. This overlap plays an important role in determining whether an external drive that induces transition between anti-ground states in different $P$ sectors  also changes the occupancy of the $b_2$ mode. To obtain it, we represent $b_2$ as the differential operator
\begin{equation}
  \label{eq:b_2_differential}
  b_2=\frac1{\sqrt2}\left(x+\frac{d}{dx}\right)
\end{equation}
in the space of functions $\psi(x)$ of a real variable $x$.  It is easy
to verify that the commutator $[b_2^{},b_2^\dagger]=1$.  The vacuum
state $\psi_0(x)$ of the operator $b$ is then defined by
\[
  \left[\frac1{\sqrt2}\left(x+\frac{d}{dx}\right)+\lambda\right]\psi_0(x)=0.
\]
This first order differential equation is easily solved,
\begin{equation}
  \label{eq:vacuum}
  \psi_0(x)=\frac{1}{\sqrt[4]\pi}\exp\left(-\frac12(x+\sqrt2\lambda)^2\right).
\end{equation}
Here we applied the normalization condition $\langle\psi|\psi\rangle=1$, where
\[
  \langle \phi|\psi\rangle=\int_{-\infty}^\infty
  \phi^*(x)\psi(x)dx.
\]
The vacuum state $\phi_0(x)$ of the operator $b_2$ is obtained by
setting $\lambda=0$ in Eq.~(\ref{eq:vacuum}).  The overlap
$\langle \phi_0 | \psi_0 \rangle$ is then obtained as
\begin{equation}
  \label{eq:b_b2_overlap}
  \langle \phi_0 | \psi_0 \rangle=e^{-\lambda^2/2}
  =\exp\left[-\frac{1}{16\ln^22}\left(\frac{P}{V_0}\right)^2\right].
\end{equation}
This result will be used in Section \ref{sec:observe}.

\subsection{Upper and lower edges of the spectrum at $P\gg1$}
\label{sec:crossover}

Next, we summarize the overall behavior at the spectral edges in the
full range of interactions.  At $P\gg1$ the lower edge of the spectrum
for weak interactions is given by
\begin{equation}
  \label{eq:lower_bound_F}
  E^{(l)}_F=-\frac{P^2}{2}-V_0P(\ln P-1)-\frac{7\pi^4}{360}V_0^2,
\end{equation}
where we combined our earlier results (\ref{eq:E_F}),
(\ref{eq:delta_E_F_1__both_Coulomb}), and
(\ref{eq:delta_E_F_2_l_Coulomb}).  For strong interactions, the lower
bound is obtained by combining Eqs.~(\ref{eq:E_B}) and
(\ref{eq:delta_E_B^l_result}),
\begin{equation}
  \label{eq:lower_bound_B}
  E^{(l)}_B=-V_0P\ln P-\gamma(\infty)\frac{P^3}{V_0}.
\end{equation}
The term $-V_0P\ln P$ appears at both small and large $V_0$.  The
remaining terms in both Eqs.~(\ref{eq:lower_bound_F}) and
(\ref{eq:lower_bound_B}) are of the same order of magnitude at
$V_0\sim P$.  One should therefore expect a smooth crossover between
the two regimes at $V_0\sim P$.

For the upper spectral edge, at weak
interactions we have
\begin{equation}
  \label{eq:upper_bound_F}
  E^{(u)}_F=\frac{P^2}{2}-V_0P(\ln P-1)+\frac{7\pi^4}{360}V_0^2,
\end{equation}
where we combined the results (\ref{eq:E_F}),
(\ref{eq:delta_E_F_1__both_Coulomb}), and
(\ref{eq:delta_E_F_2_Coulomb}).  For strong interactions, to leading
order in $1/V_0$, we have
\begin{equation}
  \label{eq:upper_bound_B}
  E^{(u)}_B=\frac{P^2}{4\ln2\, V_0},
\end{equation}
cf.~Eqs.~(\ref{eq:E_B^u_second-order}) and (\ref{eq:E_B^u}).  We
omitted the corrections to the leading contribution
(\ref{eq:upper_bound_B}) found in Eq.~(\ref{eq:E_B^u_fourth-order}),
which limits the applicability of Eq.~(\ref{eq:upper_bound_B}) to
$V_0\gg \sqrt P$.

Let us consider the regime of moderately strong interactions,
$V_0\sim V^*$, where
\begin{equation}
  \label{eq:V^*}
  V^*=\frac{P}{2\ln P}.
\end{equation}
For such interactions the last term of
Eq.~(\ref{eq:upper_bound_F}) is negligible, while the first two terms
yield $E_F^{(u)}\sim P^2$.  On the other hand,
Eq.~(\ref{eq:upper_bound_B}) yields $E_B^{(u)}\sim P\ln P\ll P^2$.
At the energy scales of the order of $P^2$ one can then approximate
$E_B^{(u)}=0$.  This gives the true upper bound of the
spectrum at sufficiently large interactions, where $E_F^{(u)}$ becomes negative.
Summarizing the above findings, at $P\gg1$ we obtain
\begin{equation}
  \label{eq:upper_bound_large_p}
  E^{(u)}=\frac{P^2}{2}\times
  \left\{
  \begin{array}[c]{ll}
    1-\frac{V_0}{V^*}, & \mbox{at $V_0<V^*$},\\[2ex]
    0, & \mbox{at $V_0> V^*$}.
  \end{array}
  \right.
\end{equation}
It is important to note that unlike the smooth crossover between the
regimes of weak and strong interactions at the lower edge of the
spectrum, at the upper one we have a sharp transition at $V_0=V^*$.

\subsection{Scar states near the upper spectral edge}
\label{sec:scars}

Interacting one dimensional Fermi systems can be described in two
complementary languages.  At weak interactions the original theory in
terms of weakly interacting fermions is more convenient.  At strong
interactions, bosonized description is preferable, as the bosons are
now weakly interacting.  In particular, this applies to both the
ground and anti-ground states, Sec.~\ref{sec:crossover}.  However, we
have found that the crossovers between the weakly and strongly
interacting regimes are qualitatively different at the two extremes of the spectrum.

For the ground state, the smooth crossover implies that at $V_0\sim P$
both fermions and bosons experience strong interactions, and neither
description is simpler than the other.  This is not the case for
the anti-ground state.  We found that at $V_0<V^*$ the highest energy
state is adequately described in terms of weakly interacting fermions.
Conversely, at $V_0>V^*$, the anti-ground state is described most
simply in the language of weakly interacting bosons.  The energies of
these two states are equal at $V_0=V^*$, but their nature remains very
different.  In other words, we conclude that at $P\gg1$ the two levels
cross at $V_0=V^*$ without significant level repulsion.

Our description of the bosonic anti-ground state in
Sec.~\ref{sec:upper_boson} applies as long as $V_0>\sqrt P$.  This
state has the highest energy at $V_0>V^*$, but in the parametrically
broad range of interaction strengths $\sqrt P\ll V_0<V^*$ it has lower
energy than the fermionic anti-ground state.  It is therefore embedded
in the nearly continuous spectrum of states composed of weakly
interacting particle-hole excitations.  This behavior is analogous to
that of scar states in nearly ergodic systems \cite{ChandranMotrunich20}.

One can similarly argue that at $V_0>V^*$ the fermionic anti-ground
state enters the region of energies dominated by states composed of
weakly interacting bosons.  Given that the structure of the
perturbation theory in small $V_0$ is the same for the ground and
anti-ground states, we expect that the result (\ref{eq:upper_bound_F})
also applies as long as $V_0\ll P$.  Thus the fermionic scar state
near the upper bound of the spectrum should be present in the
logarithmically broad range $V^*< V_0\ll P$.

The qualitative difference between the crossovers for the ground and
anti-ground states can be traced  back to their form in the limit of
strong interactions, see Eq.~(\ref{eq:psi_B}).  The overlap of the
single boson excitation state $\psi_B^{(l)}$ with $\psi_F^{(l)}$ given
by Eq.~(\ref{eq:psi_F}) is $1/\sqrt P$.  [This can be seen immediately
from Eq.~(\ref{eq:b_q}).]  On the other hand, the state $\psi_B^{(u)}$
is very different in that a single bosonic state is filled with
$P\gg1$ particles.  One can show that its overlap with $\psi_F^{(u)}$
is exponentially small at large $P$, which results in level crossing
at the upper edge of the spectrum.  This argument can be extended to
at least a few highest energy states, which should also cross each
other near $V^*$ at sufficiently large $P$.  Thus one should expect a
sequence of bosonic scar states at $V_0<V^*$ as well as several
fermionic states at $V_0>V^*$.  This conclusion is supported by
numerical calculations, see Sec.~\ref{sec:numerics}.

\section{Numerical results}
\label{sec:numerics}

\begin{figure}
\includegraphics[width=.95\columnwidth]{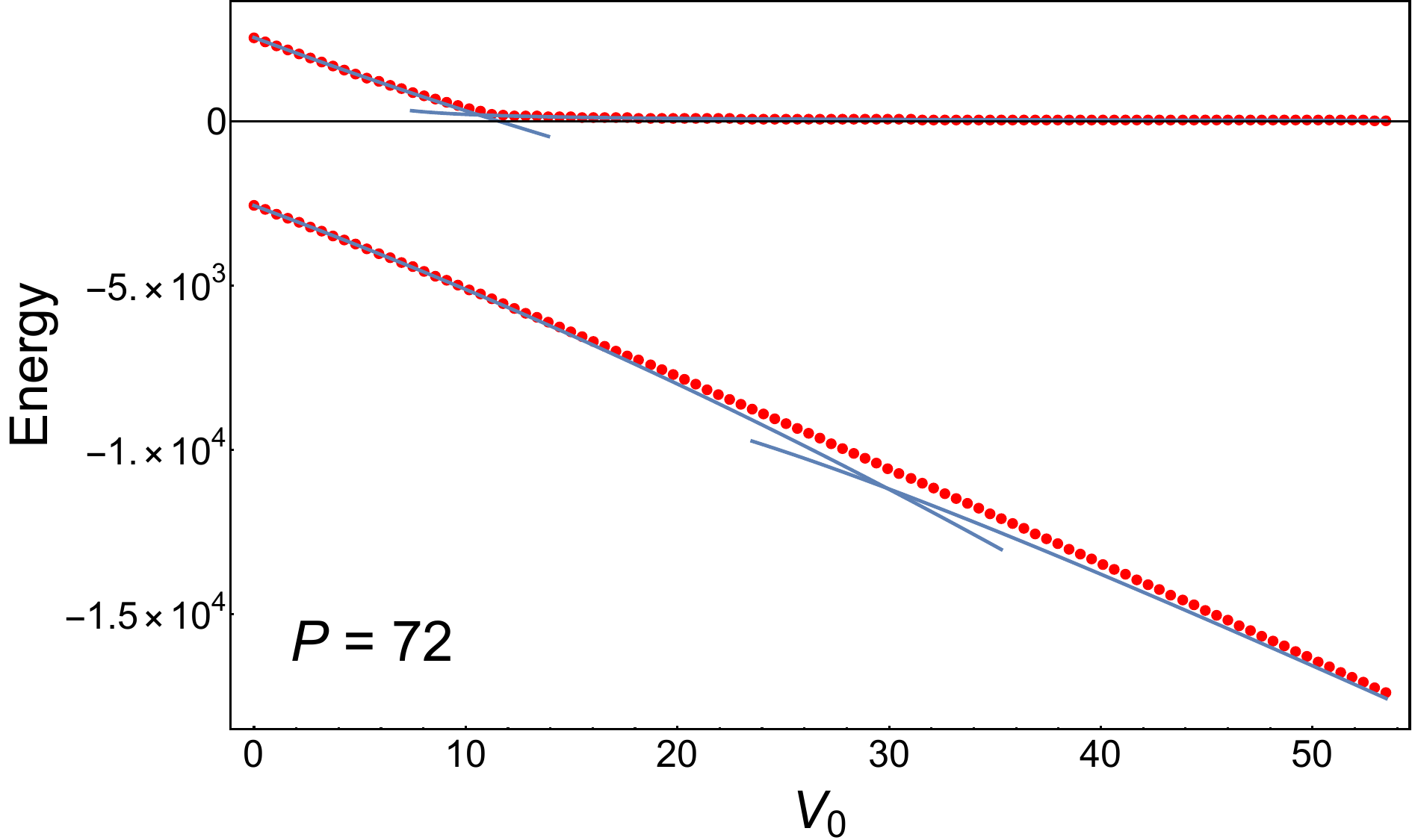}

\caption{\label{fig:HL} Upper and lower edges of the spectrum for $P=72$
  obtained by numerical diagonalization of the Hamiltonian given by Eqs.~(\ref{eq:H_B_b}) and (\ref{eq:H_F_b}), assuming Coulomb interaction (\ref{eq:V_q_Coulomb}).
  Solid lines are the results of analytical perturbation theory, see
  Eqs.~(\ref{eq:E_F}), (\ref{eq:E_B}), (\ref{eq:delta_E_F_1_Coulomb}),
  (\ref{eq:delta_E_F_1_l_Coulomb}), (\ref{eq:delta_E_F_2_Coulomb}),
  (\ref{eq:delta_E_F_2_l_Coulomb}), (\ref{eq:delta_E_B^l_result}), and
  (\ref{eq:E_B^u_fourth-order}).  Note that at the lower edge of the
  spectrum, the crossover is at larger values of $V_0$ than at
  the upper edge.}
\end{figure}

In this section we supplement the analytical results derived in the
previous sections with numerical calculations. To start off,
Fig.~\ref{fig:HL}, shows the general comparison between the
numerically obtained upper and lower edges of the spectrum and the
analytical strong and weak-coupling perturbative expansions for
$P = 72$. The agreement is excellent, and the intersections of the
analytical asymptotes tracks qualitative changes in the spectrum.  The
most notable feature is the incipient level crossing at the upper edge
of the spectrum, which is indicative of QMBS. In the following
subsections we carefully examine these level crossings and their
dependence on the form of interaction.   We
use  participation ratio in fermionic and bosonic bases to
classify the states according to their type.  This reveals that the scar
states are not limited to the upper edge of the spectrum, but exist
throughout, in the intermediate coupling regime. 

\subsection{Level crossings}

\begin{figure}
\includegraphics[width=.95\columnwidth]{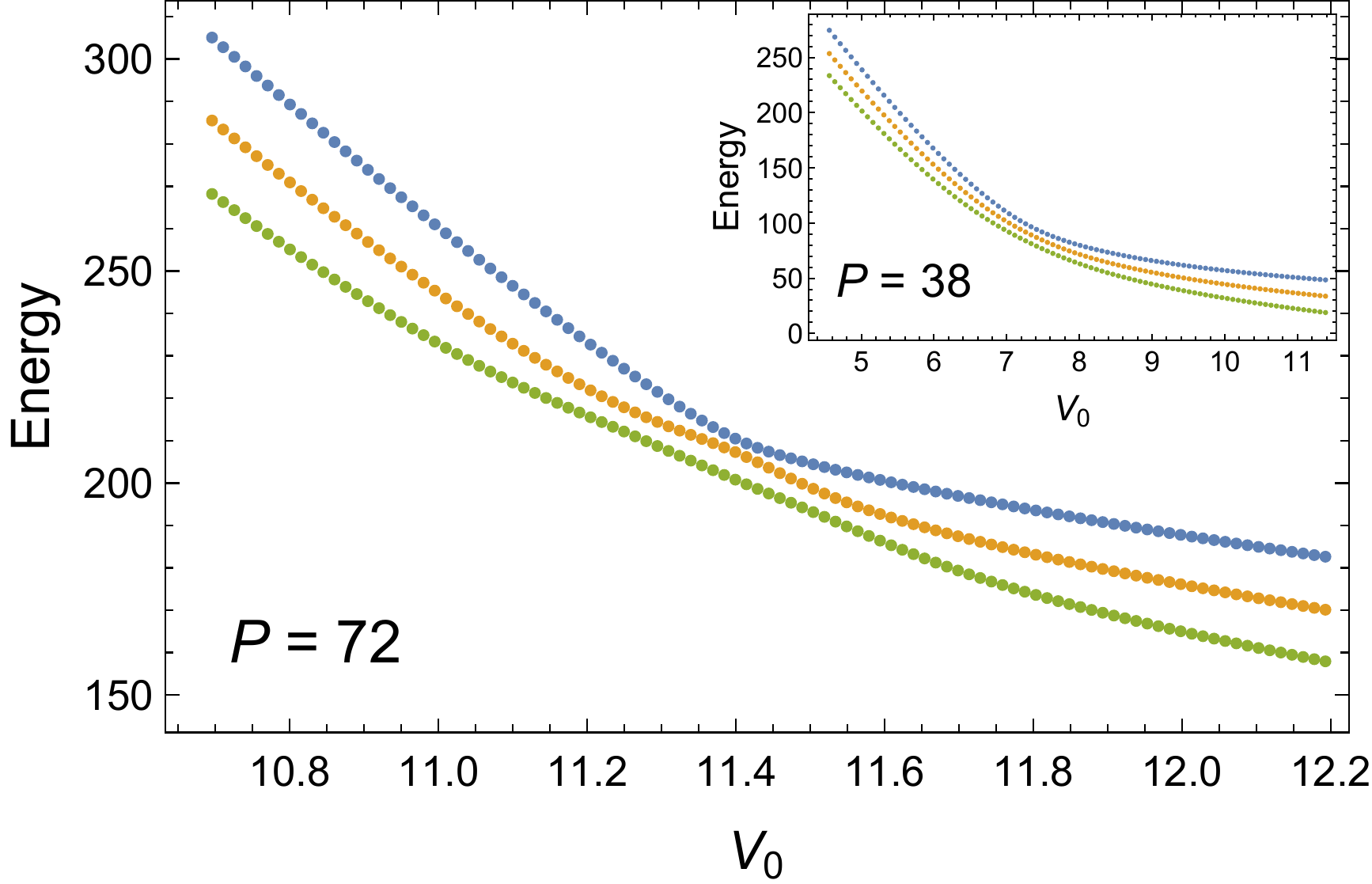}
\caption{\label{fig:3levelsC} Coulomb interaction. Three uppermost levels as a function of the interaction strength. Main panel: total momentum sector $P = 72$. Note that the second level (orange dots) has regions of positive and negative curvature, separated by two inflection points. Less obviously, third level (green dots) has four inflection points. This behavior is reflective of avoided level crossings. Inset: total momentum sector $P = 38$. The data shows no avoided crossing behavior.
}
\end{figure}

As shown analytically above, for repulsive interactions, we anticipate
a level crossing between the fermionic anti-ground state
originating at low interactions strengths and bosonic anti-ground
state, which becomes exact in the infinite interaction limit. The 
crossing as a function of the interaction scale $V_0$ at fixed $P$ 
is expected to develop only as $P\to \infty$. In
Fig.~\ref{fig:3levelsC} we show three uppermost levels in the spectrum
of Coulomb-interacting fermions for $P = 72$. Their most noticeable
feature is that levels congregate near interaction $V_0 = 11.4$.  Upon
closer inspection, the changes of curvature reveal that the levels
undergo a pair-wise avoided level crossings.  At a smaller value of
the total momentum $P=38$ (inset) no avoided crossing is evident.

\begin{figure}
\includegraphics[width=.95\columnwidth]{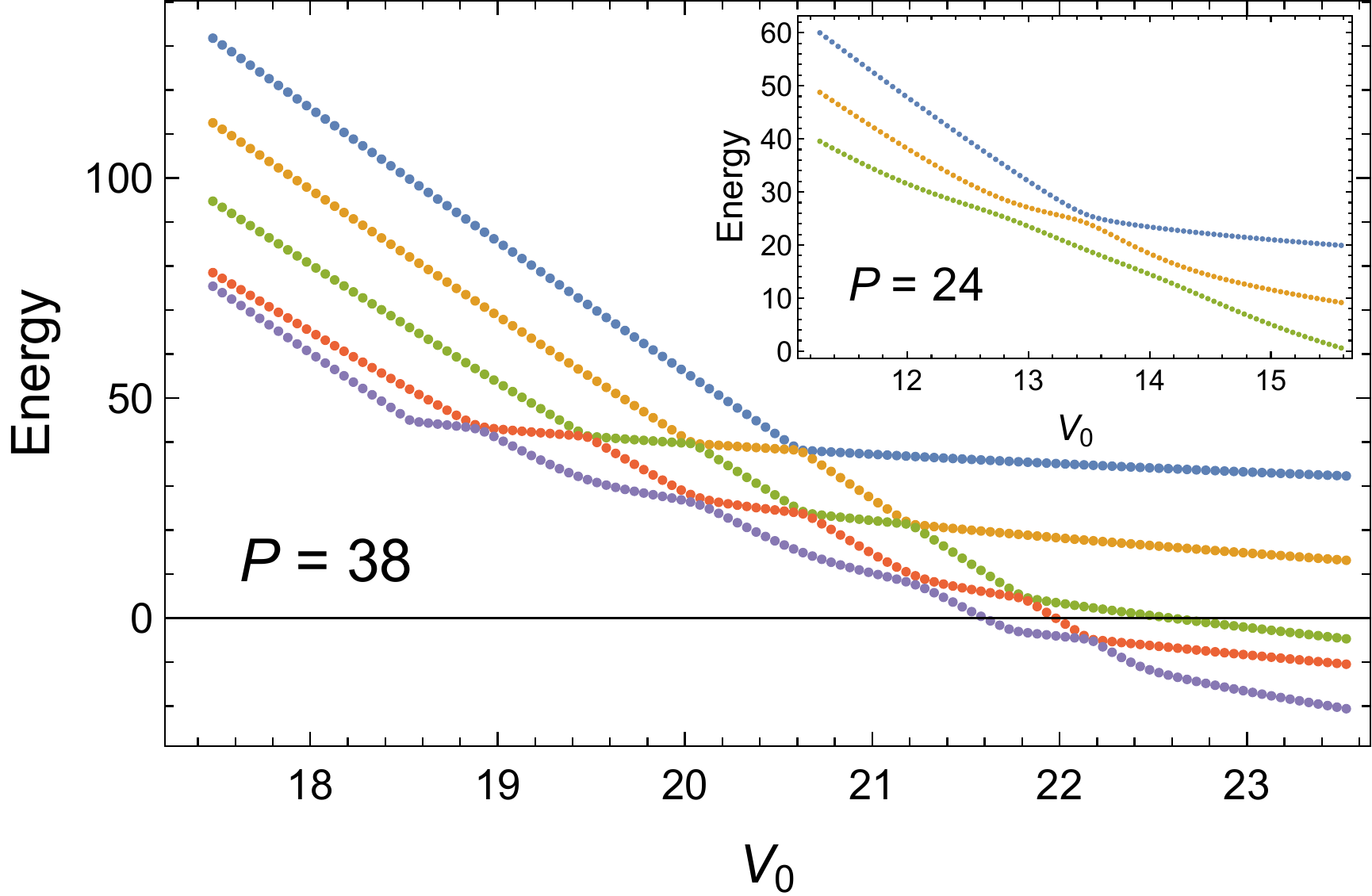}\\
\caption{\label{fig:3levelsLog} Logarithmic interaction. Uppermost levels as a function of interaction strength.  Main panel: total momentum sector $P = 38$. Compared to Fig.~\ref{fig:3levelsC}, multiple level crossings have become apparent, even at a considerably smaller value of $P$. The level repulsion is discernible only between 4th and 5th levels, or for smaller values of the total momentum, such as $P = 24$ shown in the inset. 
}
\end{figure}

Though the minigaps of the avoided level crossing clearly shrink with increasing $P$, our numerical limitations do not allow us to readily access larger system sizes to see the incipient level crossing more explicitly in the case of Coulomb interaction. 
Fortunately, it is possible to make the underlying physics more self-evident by modifying the form of interaction. 
In Fig.~\ref{fig:3levelsLog} we used interaction $V(q)=V_0(|q|^{-1}-1)$ that scales as $\ln\frac{1}{|x|}$ in real space.  (It can be viewed as a special case of the more general interaction $V(q)=V_0(|q|^{\alpha-1}-1)/(1-\alpha)$ corresponding to $\alpha=0$, whereas the Coulomb interaction (\ref{eq:V_q_Coulomb}) corresponds to the limit $\alpha \to 1$). Already at the moderate system size $P = 38$, the minigaps become much smaller than the level spacing for the top levels, increasing gradually for deeper levels. For comparison, the inset shows the data for $P=24$, which has a level structure similar to that for a much larger value $P = 72$ and Coulomb interactions.

Despite the significant quantitative difference the structure of the energy levels near the transition point for the Coulomb and logarithmic interactions, we observe no qualitative changes in the physical behavior.  Given that the modified interaction is much more amendable to the numerical exploration, we limit ourselves to this case in the rest of this Section.

The level crossings are a strong indicator that some states can
penetrate into the interaction regime dominated by states of different
character. That is, they point to the presence of scars.  Next, we verify that this is indeed the case by directly examining the character of the wavefunctions.

\subsection{Wavefunction character}

The presence of scars is evidenced not only by the energy level
crossings, but also by the character of the eignestates
$|\Psi_m\rangle$. To demonstrate that, we expand the exact eigenstates
in terms of the noninteracting (fermionic) basis states $|f_n\rangle$
or infinitely interacting (bosonic) basis states $ |b_n\rangle$. We
then quantify the number of the relevant states in the expansion using
the corresponding participation ratios,
$N_f(m) = 1/\sum_n |\langle\Psi_m|f_n\rangle|^2$ and
$N_b(m) = 1/\sum_n |\langle\Psi_m|b_n\rangle|^2$. For states
accurately represented by only a few basis states the corresponding
$N$ is of order unity.  On the other hand, if the full basis is needed
to represent a state, then $N$ scales with the dimensionality of the
Hilbert space.  It is convenient to also introduce a quantity
$\chi(m) = N_b(m)/N_f(m)$.

\begin{figure}
\includegraphics[width=.95\columnwidth]{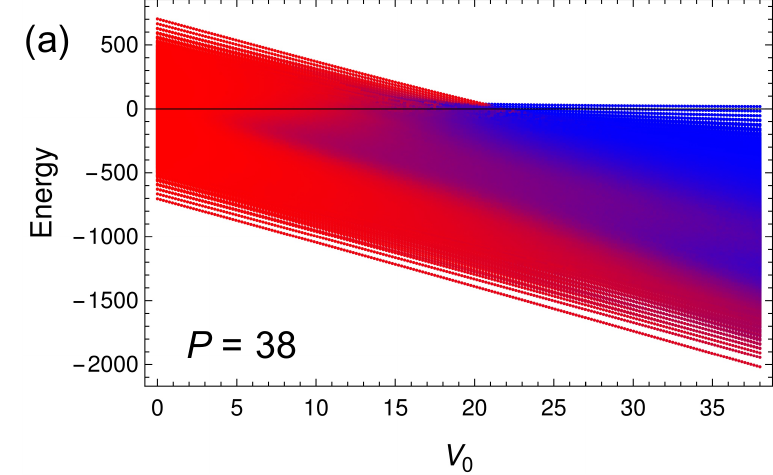}\\[3ex]
\includegraphics[width=.95\columnwidth]{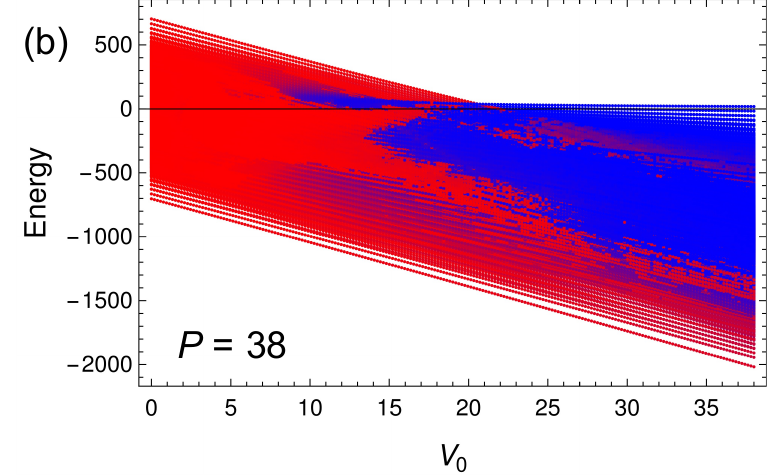}\\

\caption{\label{fig:PR} Character of eigenstates, as encoded by the
  ratio of fermionic and bosonic participation ratios. Red corresponds
  to fermionic character, and blue to bosonic. Panel (a) shows typical
  character.  Panel (b) emphasizes the outliers---the states that most
  strongly deviate from the typical states in their vicinity. (See
  text for details.)  Most notably, near the upper bound of the
  spectrum, there is an island of atypical bosonic scar states that
  appears at low interactions ($V_0\sim 12$) and a fermionic island at
  large interactions ($V_0\sim 25$).  Both appear to be the
  extrapolations of the corresponding states at the upper edge of the
  spectrum.  }
\end{figure}

In Fig.~\ref{fig:PR} we show the spectrum of the system with
logarithmic interaction in the sector $P = 38$ as a function of the
interaction strength, with each dot representing a state. The color
indicates the character of the state: red corresponds to the fermionic
character, $\chi(m) \gg 1$, blue to bosonic, $\chi(m) \ll 1$, and
intermediate colors correspond to intermediate character. Since the
number of states at this value of $P$ is very large (26015), the order
in which the states are plotted matters. There are two particularly
informative methods, which highlight the most typical or the least
typical states. We first compute the deviation of $\chi(m)$ from the
geometric mean $\bar\chi(m)$ of its 100 neighbors in energy (on both
sides, at the same $V_0$).  In the top panel, the typical states, for
which $\chi(m)$ is closest to $\bar\chi(m)$ are plotted last.  This
naturally leads to a smooth pattern.  In the lower panel, the outliers
are plotted last, which emphasizes the atypical, scar states.

As was theoretically anticipated, bosonic-like states near the upper
edge of the spectrum penetrate into the fermionic domain.
Similarly, fermionic-like states can appear on the strongly coupled
side that is dominated by bosonic states.  Notably, there are also a
number of additional scar states throughout the spectrum.
More generally, the participation ratio shows that in the transition
region, the fermionic and bosonic states intermix in a complicated
fashion throughout the spectrum, even in a fixed total momentum
sector.

\section{Observing scars}
\label{sec:observe}

In the previous sections we have shown that for a sufficiently large value of total momentum $P$ atypical scar states appear throughout the spectrum, most prominently in the vicinity of the upper edge. The discontinuity in slope as a function of the interaction strength indicates that for smaller interactions the  states are fermionic, and for larger---bosonic in character.  We now address observability of the atypical bosonic collective states in the regime of intermediate and weak interactions, where the  native states are fermionic.

\begin{figure}
\includegraphics[width=.9\columnwidth]{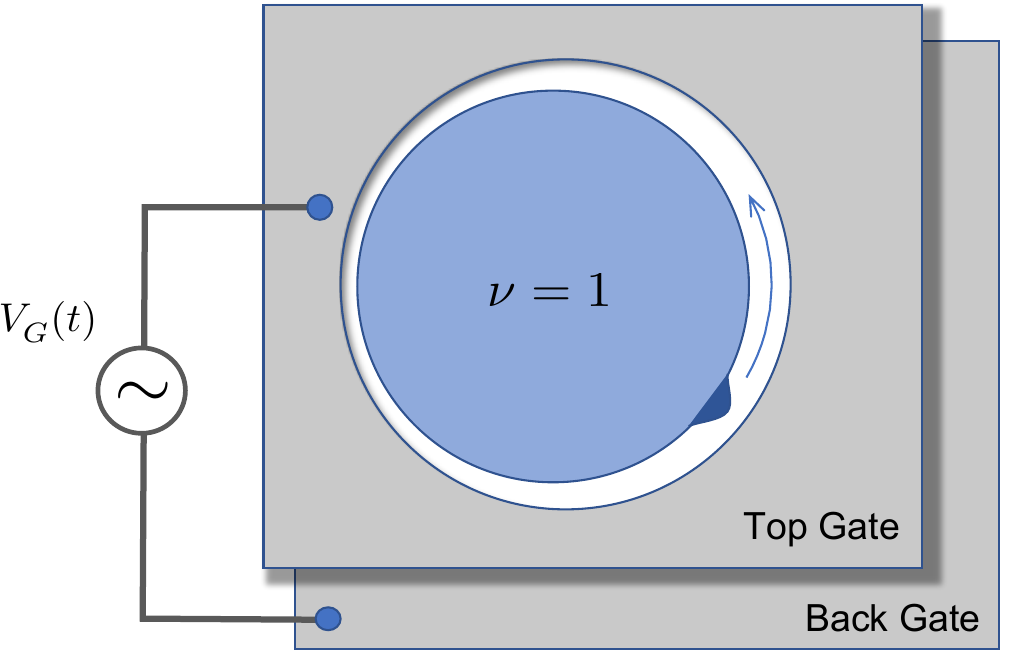}
\caption{\label{fig:setup} Interacting chiral fermions  at the edge of a $\nu = 1$ quantum Hall droplet. Starting from the ground state, $P = 0$, atypical scar states at the upper edge of the spectrum can be excited by applying voltage $V(t)$ between the top and the back gates at the frequency that corresponds to a wavepacket (blue bump) propagation around the edge. This favors transitions between $P$ and $P+1$ anti-ground states. Selective excitation of these states is further enhanced by having the gate primarily coupled to the lowest spatial harmonic (for circular droplet shown here, the top gate is slightly offset relative to it).
Once exited, voltage oscillations associated with a travelling density wave can persist for anomalously long time.    
}
\end{figure}

As a specific practical realization we consider a quantum Hall
droplet, electrically insulated from the rest of the system, but
electrostatically gated. Voltage applied between back gate and the top gate can be used to excite density modes in the droplet, and the
top gate can be also used to detect excited oscillations,
Fig.~\ref{fig:setup}.    We assume for
simplicity that the dominant spatial dependence of the gate potential
is harmonic with the period $L$ (as, for instance, would be the case
for a circular droplet and circular gate opening surrounding it, but
slightly displaced). Then, the Hamiltonian that describes the
interaction between the gate and the chiral system is
\begin{equation}\label{eq:dr}
H_D = D \cos (\omega_D t) (b_1 + b_1^\dag),
\end{equation}
where $D$ is the drive strength and $\omega_D$ is the drive frequency.  From Eq.~(\ref{eq:E_B^u_fourth-order}), the upper edge of the spectrum, from the bosonic side is
\begin{equation}\label{eq:EU}
    E_B^u(P)= \frac{P(P-1)}{4\ln 2\, V_0 } + O(P^3/V_0^3).
\end{equation}
As explained above, these correspond to the multiboson states, $\Psi_P \approx (b_1^\dag)^P|0\rangle$.  The drive Eq.~(\ref{eq:dr}) effectively acts as hopping on a one-dimensional lattice with sites labeled by $P$. If we assume that the drive frequency is resonant with the $P = 0 $ to $P = 1$ transition, it will connect states up to 
\begin{equation}\label{eq:splitting}
    |E_B^u(P_{max})- E_B^u(0)| \lesssim \sqrt{P_{max}} D,
\end{equation}
where the pefactor $\sqrt{P_{max}}$ on the right hand side is the scale of $|\langle b_1 \rangle|$. This  implies that applying drive  of strength $D$ leads to excitation of multiboson anti-ground states up to 
\begin{equation}\label{eq:pmax}
P_{max} (D) \sim (D V_0 )^{2/3}. 
\end{equation}
From Eq.~(\ref{eq:pmax}), the drive strength needed to reach the kink ($V_0 \sim P$)  at the upper spectral edge is $D_{\rm kink} \sim V_0^{1/2}$.  

The above assumes that the $b_2$ bosons are not being excited. We now verify that this is indeed the case. Referring back to Sec.~\ref{sec:upper_boson} and in particular the effective Hamiltonian (\ref{eq:H_eff}), we can estimate the matrix element of the transition to states with non-zero  number of excited phonons $b_2^\dag b_2^{}$.  From Eq.~(\ref{eq:b_b2_overlap}), the probability to remain in the vacuum of $b_2$ while changing $P$ to $P+1$ is given by the Frank-Condon factor, $e^{- 1/({8\ln^22\,V_0^2})}$, which is close to 1 for large $V_0$ that we assume here. After $P$ consecutive steps, starting from $P = 0$ state,  this probability  becomes $e^{- P/({8\ln^22\,V_0^2})}$, remaining close to one as long as $P \ll V_0^2$.  This is indeed the regime where the perturbation theory developed in Sec.~\ref{sec:upper_boson} is expected to hold. (Recall that the bosonic ``scar regime" is approximately $\sqrt{P}< V_0<P$.)
We thus conclude that the bosonic scar anti-ground state can be effectively excited, and should be observable via voltage oscillations on the gate long after the drive has been turned off.

\section{Discussion and open questions}
\label{sec:discussion}

In this paper we have found evidence of quantum many-body scars in a simple model of one-dimensional chiral interacting fermions. The evidence came from  level crossings in the spectrum and analysis of the character of the wavefunctions.

Typically, level crossings are a consequence of a symmetry or integrability of a system. However, in the case we study there, there is no symmetry that protects the crossings that we observe, since they develop rapidly with increasing $P$, while no new symmetries are generated.   As a practical method to rule out integrability, we employ the so-called $r$-statistics  \cite{wigner1993characteristic, PhysRevB.75.155111, Atas2013}, which is commonly used to distinguish between integrable and chaotic systems. From the eigenvalue spectrum $E_i$, we compute the average $r =  \langle \min(\delta_i, \delta_{i-1})/\max(\delta_i, \delta_{i-1})\rangle$, where $\delta_i = E_{i+1} - E_{i}$. For uncorrelated random levels (no level repulsion, Poisson statistics) this value is $r_P\approx  0.386$; for a chaotic system with time reversal symmetry, the level distribution is expected to follow the Gaussian Orthogonal Ensemble (GOE) of Wigner-Dyson statistics with $r_{\rm GOE} \approx 0.53$.
\footnote{Note that even though quantum Hall systems break time reversal invariance, the one-dimensional edge state in the presence of translation symmetry can be fully described by a real Hamiltonian Eqs.~(\ref{eq:H_F_f}) and (\ref{eq:H_B_f}).  This makes comparison with GOE relevant for our system.}

\begin{figure}
\includegraphics[width=.95\columnwidth]{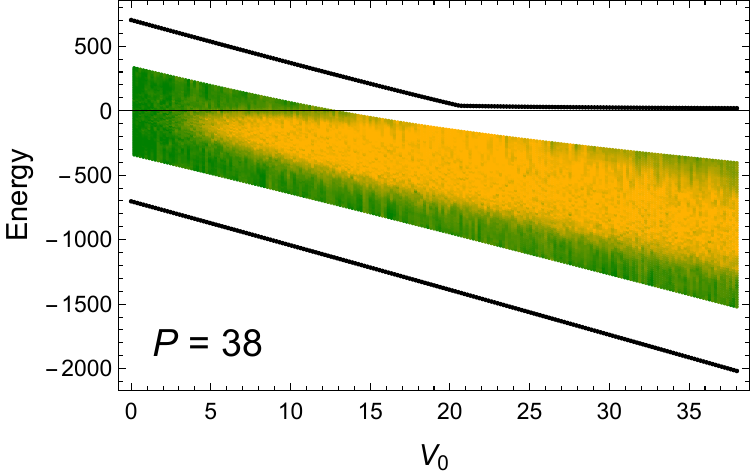}\\
\caption{\label{fig:wignerness} Energy-resolved r-statistic as a function of interaction strength, log-interaction, $P = 38$. Both the energy and $r$ are averaged over 1001 consecutive states in the middle of the spectrum, and no fewer than 501 states near the edges.
Orange regions correspond to Wigner-Dyson, while green to Poisson statistics. Thick black lines show the positions of the upper and lower bounds of the spectrum.
}
\end{figure}

The full-spectrum average of $r$ confirms that the system is non-integrable away from the zero and infinite interaction limits.
This confirms that the level crossings that develop near the upper edge of the spectrum (for fixed large $P$) are not a consequence of integrability. We refine this analysis by adding energy resolution, which provides more detailed information about the way ergodicity is reached in different parts of the spectrum, see Fig.~\ref{fig:wignerness}.  Starting from the small interactions, where  $r(\bar E) 
\approx r_{P}$, at intermediate $V_0$ we see that $r(\bar E)$ 
approaches $r_{GOE}$, first in the middle of the spectrum, and then throughout. 

Interestingly, the upper boundary that separates the Wigner-Dyson from Poisson level statistics approximately matches the extrapolation of the upper edge of the spectrum from high to low interactions, and vice versa. 
This is also where the fermionic and bosonic islands of scar states reside, see Fig.~\ref{fig:PR}(b).

In addition to the QMBS near the top of the spectrum, there are other atypical states {\em in the middle\/} of the spectrum that are prominently present in Fig.~\ref{fig:PR}(b) at intermediate couplings. 
These states have very large deviation from their typical neighbors in terms of the participation ratios. It is also notable that these atypical states form regular patterns in the energy-interaction strength plane, extending far beyond their parent domains.  
It thus appears that they also are a result of persistent level crossings, similar to the situation near the upper edge of the spectrum.
We leave the determination of the physical nature of these states and their number  for future work.  

Similar physics may also arise in other models. Unfortunately, the most direct extension to non-chiral fermions makes the Hilbert space infinite in every total momentum sector $P$, making the numerical analysis more complex. A rich variety of quantum spin models with and without quantum phase transitions may provide an alternative  playground to test the connection between integrability and QMBS.  In combination, these studies may shed light on the question of how general or particular QMBS are in many body systems, and their role in quantum dynamics and thermalization.

\begin{acknowledgments}
This work was supported by the US Department of Energy, Office of Science, Materials Sciences and Engineering Division.
\end{acknowledgments}

\bibliography{refs}

\begin{thebibliography}{40}
\bibitem [{one()}]{onefootnote}%
  \BibitemOpen
  \href@noop {} {}\bibinfo {note} {The expressions for $c(\infty)$ and
  $\gamma(\infty)$ quoted below Eqs.~(\ref{eq:delta_E_F_2_l_Coulomb}) and
  (\ref{eq:gamma}) were obtained by converting the sums in Eqs.~(\ref{eq:c(p)})
  and (\ref{eq:gamma}) to integrals.}\BibitemShut {Stop}%
\end{thebibliography}

\end{document}